\def\epsfpreprint{Y}   
                       
                       \if \epsfpreprint Y
\documentstyle[12pt,epsf]{article} \fi \if \epsfpreprint N
\documentstyle[12pt]{article} \fi

\def\figsize{7.2}

\def\figure#1#2#3{\if \epsfpreprint Y \epsfxsize=#3truein
\centerline{\epsffile{fig_#1.eps}}
\centerline{\vbox{{\bf \noindent Figure #1.} #2}}
\bigskip \fi}

\def\qbarq{\langle \overline{q} q \rangle}

\def\spose#1{\hbox to 0pt{#1\hss}}
\def\ltapprox{\mathrel{\spose{\lower 3pt\hbox{$\mathchar"218$}}
 \raise 2.0pt\hbox{$\mathchar"13C$}}}
\def\gtapprox{\mathrel{\spose{\lower 3pt\hbox{$\mathchar"218$}}
 \raise 2.0pt\hbox{$\mathchar"13E$}}}
\def\inapprox{\mathrel{\spose{\lower 3pt\hbox{$\mathchar"218$}}
 \raise 2.0pt\hbox{$\mathchar"232$}}}

\def\one{The expected functional form of $\qbarq$ vs.\ $m_f$. 
The solid line corresponds to staggered fermions and 
the dashed line to domain wall fermions.}
\def\two{$\qbarq$ vs.\ $m_f$ for staggered fermions. The left
columns are for an $8^4$ volume and the right columns for a
$16^4$ volume. The $\zeta=0$ figures correspond to the
compactified singular-gauge instanton background. The 
$\zeta=0.01, 0.1$ figures correspond to the same 
background but random noise has been superimposed with 
amplitude $\zeta$.}
\def\three{Same as Figure $2$ but for domain wall fermions at $m_0=1.2$.
Four different values of $L_s$ are shown with the circles, 
squares, crosses, and diamonds corresponding to 
$L_s=4, 6, 8, 10$.}
\def\four{Same as the $16^4$ lattice, $L_s=10$, $\zeta=0.1$
plot of Figure $3$. The solid line is a fit to $c_{-1}/m_f$
for $10^{-5} \leq m_f \leq 10^{-3}$.
The fit has a $\chi^2$ per degree of freedom $\approx 0.5$.}
\def\five{Same as Figure $3$ but now vs.\ $m_0$ at fixed 
$m_f = 5 \times 10^{-4}$.
The index of the Dirac operator changes at the vertical lines
from zero to one and then back from one to zero.
Four different values of $L_s$ are shown with the circles, 
squares, crosses, and diamonds corresponding to 
$L_s=4, 6, 8, 10$.}
\def\six{Same as Figure $3$ but for $m_0=0.75$. The circles, squares, 
crosses, diamonds and plus symbols correspond to $L_s = 4, 6, 8, 10,
12$ respectively. The star symbols on the $8^4$ graph correspond to
$L_s=24$.}
\def\seven{Same as Figure $6$ but for $m_0=0.5$ for the $8^4$ and 
$m_0=0.25$ for the $16^4$. The circles, squares, crosses and diamond
symbols correspond to $L_s = 8, 12, 16, 24$ respectively. The plus
symbols on the $8^4$ graph correspond to $L_s=48$.}
\def\eight{$\qbarq$ vs. $L_s$ for zero noise amplitude, 
fixed $m_f = 5 \times 10^{-4}$ and for three values of $m_0 = 0.5,
0.75$ and $1.2$ corresponding to circles, squares and crosses. The
fits are to a function of the form $1 / [ c_0 + c_1 e^{-c_2
L_s}]$. For the three values of $m_0$ the fitting range of $L_s$ is
$[8,48]$, $[6,48]$ and $[4,48]$, the coefficient $c_0 = 0.183(8),
0.44(3)$ and $0.91(31)$, and the $\chi^2/{\rm dof}$ is $4 \times
10^{-2}$, $4 \times 10^{-2}$ and $2 \times 10^{-4}$.}
\begin{document}

\title{\bf Domain wall fermion zero modes on classical topological backgrounds}
\vskip 1. truein
\author{
P. Chen, N. Christ, G. Fleming, A. Kaehler, \\
C.  Malureanu, R. Mawhinney, C. Sui, P. Vranas and Y. Zhestkov \\
\\
Columbia University \\ 
Physics Department \\ 
New York, NY 10027 }
\maketitle

\vskip -4.0 truein
\rightline{CU-TP-906}
\vskip 5.3 truein

\begin{abstract}

The domain wall approach to lattice fermions employs an
additional dimension, in which gauge fields are merely replicated, to
separate the chiral components of a Dirac fermion.  It is known that in
the limit of infinite separation in this new dimension, domain wall
fermions have exact zero modes, even for gauge fields which are not
smooth. We explore the effects of finite extent in the fifth dimension
on the zero modes for both smooth and non-smooth topological
configurations and find that a fifth dimension of around ten sites is
sufficient to clearly show zero mode effects. This small value for the
extent of the fifth dimension indicates the practical utility of this
technique for numerical simulations of QCD.

\end{abstract}

\newpage

\section{Introduction}

The anomalous breaking of the flavor singlet axial symmetry of QCD,
$U_A(1)$, has
important physical consequences. It is responsible for the relatively
large mass of the $\eta^\prime$ \cite{tHooft}, \cite{Witten_Ven} and
effects the order of the finite temperature phase transition
\cite{Pizarski}. A central role in this is played
by the index theorem \cite{index} which relates the winding
of the gauge field with the number of zero modes of the Dirac operator.

On the lattice the winding of the gauge field can not be defined
unambiguously and the index of a finite-dimensional lattice Dirac operator 
is necessarily zero.  Therefore, strictly speaking, the index
theorem is not valid on the lattice. However, an approximate form of
this theorem may still be present and govern anomalous effects.
Some time ago this issue was investigated in \cite{Smit_Vink} for the
case of staggered and Wilson fermions.  A first step in that direction
was to ask to what extent the lattice Dirac operator for staggered and
Wilson fermions develops the appropriate number of zero modes for a
fixed smooth gauge field background with given winding. It was found that 
for a particular background the
appropriate number of zero modes was generated but they were not
robust when high frequency noise was superimposed. Furthermore, the
background chosen was of a very special
nature and did not have the spatial variation present, for example, in an
instanton-like background.  Recent work \cite{Kaehler}, with
more realistic instanton-like backgrounds, has demonstrated that
staggered fermions do not develop zero modes unless the lattice
spacing, $a$, becomes very small ($a/D \ll 0.1$, where $D$ is the
instanton diameter).  These difficulties suggest that for QCD, staggered 
fermions may fail to reproduce anomalous effects unless the lattice 
spacing is made quite small---considerably smaller than that used in present 
QCD thermodynamics studies, performed with $N_T \le 8$ and staggered fermions.

Current numerical results for the order of the finite temperature phase
transition for two flavor QCD with light quarks show that it is not
first order \cite{order_trans}.  This is consistent with the analysis in
\cite{Pizarski}, provided $U_A(1)$
is broken.  However, current lattice discretizations manifestly
break the symmetry of the classical action at finite lattice spacing
and the corresponding discretized Dirac operators have difficulty
seeing zero modes for non-smooth topological configurations.
This raises the possibility that the present apparent second-order character
of the two-flavor QCD phase transition may be a result of lattice artifacts
rather than physical, anomalous symmetry breaking.
A recent explicit calculation of an anomalous difference of susceptibilities 
near the two-flavor QCD phase transition, using a spectral sum-rule 
sensitive to zero modes, showed effects at or below the 15\% 
level \cite{omega}.  However, in other work \cite{Kogut_Lagae_Sinc}, 
suggestions of anomalous zero-mode effects were seen.  Clearly, a lattice
fermion formulation whose action has the full global symmetry content
of the continuum theory at finite lattice spacing would be a very
useful tool for studying anomalous symmetry breaking effects.

\section{Domain Wall Fermions}

A novel approach with spectacular zero mode properties has been
developed during the past few years. Domain Wall Fermions were
first introduced in \cite{Kaplan} and a variant suitable for vector
gauge theories was introduced in \cite{Shamir}, \cite{Furman_Shamir},
\cite{BDF}.  (For reviews see \cite{DWF_reviews} and references therein.)
In this approach an extra space time dimension (henceforth to be
called ``$s$'') is introduced with free boundary conditions on the
two four-dimensional boundaries, $s=0$ and $s=L_s-1$, where $L_s$ is 
the extent of this new fifth dimension in lattice units. The
fermion fields are defined in this extended space-time but the gauge
fields are still defined on the ordinary space time and have no $s$
dependence and no $s$ component \cite{NN1}. In this sense, the extra
direction can be thought of as a sophisticated ``flavor'' space. Along
the $s$ direction the fermion field develops surface modes that
are exponentially bound to the two free boundaries (domain walls) with
the plus chirality component of the Dirac spinor on one boundary and
the minus chirality on the other. As the size in lattice spacings
($L_s$) of this direction is increased the two chiral components get
separated with only an exponentially small overlap remaining. For
finite $L_s$ this overlap breaks chiral symmetry by an exponentially small
amount and as $L_s$ tends to infinity chiral symmetry is
restored. Therefore, $L_s$ provides a new parameter that can be used
to control the regularization induced chiral symmetry breaking at {\it
any} lattice spacing. For the first time the approach to the chiral
limit has been separated from the approach to the continuum limit.

An appealing aspect of the domain wall fermion formulation is the
fact that the chirally symmetric, $L_s \rightarrow \infty$ limit
can be analyzed in some detail using the overlap formalism
\cite{NN1}.  Central to this overlap method are two Fock space
states, designated $|0_H\rangle$ and $|0^\prime\rangle$ in the notation of
\cite{Furman_Shamir}, constructed from four-dimensional, 
single-particle, fermionic states.  Because the gauge field is
independent of $s$, it is possible to develop an $s$-independent
transfer matrix ($T$) and associated Hamiltonian along this direction.
Now in the $L_s \rightarrow \infty$ limit, $T^{L_s}$ becomes a
projection operator onto the vacuum state, defined as $|0_H\rangle$,
the Fock-space state in which all the negative energy states of $H$
are filled.  The second state $|0^\prime\rangle$ is a much simpler,
kinematic construction in which all single-particle eigenstates of
position and $\gamma^5$, with negative $\gamma^5$ eigenvalue, are
filled.
 
A fermion Green's function for a given gauge field background can
then be expressed as a simple matrix element of the appropriate number
of creation and annihilation operators inserted between the states
$|0^\prime\rangle$ and $|0_H\rangle$.  In particular, the
five-dimensional fermion determinant in the massless case is simply proportional to
$|\langle0^\prime|0_H\rangle|^2$ which, for a finite space-time volume,
can be calculated explicitly numerically.  If the number of filled
levels in $|0_H\rangle$ and $|0^\prime\rangle$ is the same then
$|\langle0^\prime|0_H\rangle|^2 \neq 0$.  However, if these filling
levels differ, then $|\langle0^\prime|0_H\rangle|^2 = 0$ for
zero mass and finite volume. This implies the presence of exact zero
modes in the five-dimensional formulation.  For a Green's function to
be non-zero, an appropriate number of creation and annihilation
operators must be inserted to balance the deficit.  The deficit is
naturally integer-valued and is defined as the index of the lattice
Dirac operator.  Therefore, this method provides a way to associate an
index with the lattice fermion operator in the limit $L_s \rightarrow
\infty$ but at fixed lattice spacing \cite{NN1}.

More specifically it was found \cite{NN1}, \cite{EHN1} that for
classical backgrounds these zero modes are exact, that the deficit is
equal to the winding of the gauge field and that they exhibit all the
properties that are expected in the continuum.  It was also found in
\cite{NN1} that these modes are robust under the addition of high frequency noise.
Using the overlap formalism, numerical simulations of the massless
\cite{NNV} and massive \cite{PMV} vector Schwinger model gave the
expected value for the anomalously generated t'Hooft vertex
\cite{NNV}. Furthermore, the index of the Dirac operator was calculated in
SU(2) pure gauge theory slightly above the zero temperature crossover
region \cite{NV}. The index agreed within one sigma with the
topological charge as calculated in \cite{deForcrand} indicating that the
index theorem holds in a statistical sense. Similar studies were done
in \cite{EHN2} for pure SU(3) gauge theory.

The above mentioned body of work indicates that the overlap ($L_s
\rightarrow \infty$ limit of domain wall fermions) successfully incorporates exact zero modes at
finite and relatively large lattice spacings.  This suggests that
with domain wall fermions, lattice QCD describes anomalous effects with regularization
artifacts under firm control. Unfortunately, a direct implementation
of the overlap formalism in QCD needs computing resources that are
beyond the capabilities of present day supercomputers. While 
a new proposal for reducing the computational cost of the 
overlap has been made \cite{Neuberger}, \cite{EHN4}, it is not yet clear
what the QCD computing requirement of this method will be. On the
other hand, a straight-forward approach is to keep $L_s$ finite. This
method was used in \cite{PMV} to simulate the dynamical two flavor
Schwinger model. In that work a detailed analysis of the $L_s$
dependence indicated that for the massive theory the $L_s=\infty$
value of the chiral condensate and of the t'Hooft vertex was already
reproduced within a few percent at $L_s \approx 10$.  Furthermore, 
the rate of approach
was consistent with exponential decay with a decay rate that became
faster as the continuum limit was approached. Also, for an application
to quenched QCD see \cite{Blum_Soni}.

\section{Numerical Results}

These promising results indicate the important possibility of
practical, chirally-consistent QCD simulations with domain wall
fermions at finite $L_s$. In preparation for such simulations we wish
to investigate to what extent the zero mode properties of the $L_s =
\infty$ theory are maintained at numerically accessible values of
$L_s$ ($\sim 10 - 20$).  If a much larger $L_s$ is needed then one
would not be able to exploit these important features. On the other
hand, if these properties are maintained at accessible values of $L_s$
then anomalous effects can be studied with firm control over the
finite $L_s$ artifacts.

Here we investigate this question using a classical instanton-like
background \cite{Kaehler} with prescribed winding of one unit.  This
is a compactified, singular-gauge instanton with origin in the center
of a unit hypercube. The instanton field is cut off smoothly at some radius
$r_{\rm max}$ so that it is entirely contained within the lattice volume. Specifically,
we begin with:
\begin{equation}
A_{\mu}(x) = 
-i \sum_{j=1}^{3} \eta^{j \mu\nu} \lambda^j {x_{\nu} \over {x^2 + \rho(r)^2}} \ \ ,
\ \ \ \ \ \  \rho(r) = \rho_0 ( 1 - {r \over {r_{\rm max}}} ) \Theta( r_{\rm max} - r )
\label{inst}
\end{equation}
where $A_\mu$ is the gauge field potential, $x_\nu$ is the space-time
coordinate, $r$ is the magnitude of $x$, $\lambda^j, \ j=1,2,3$ are
the first three Gell-Mann matrices, $\eta^{j\mu\nu}$ is as in
\cite{tHooft}, $\Theta$ is the usual Heavyside function 
and $\rho_0$ is the instanton radius. Outside the fixed
radius $r_{\rm max}$ the configuration is strictly a gauge
transformation.  This continuum field is then transcribed in the
standard way to a lattice configuration of group elements, $U_\mu(x)$, 
defined on the lattice links $x, \mu$.
Finally, the lattice equivalent of the continuum transformation to
singular gauge is applied:
\begin{equation}
U_{\mu}(x) \rightarrow g(x)U_{\mu}(x) g^{-1}(x + a_{\mu}) \ \ , 
\ \ \ \ \ \ g(x) = \sum_{j=0}^{3} \frac{x_j \lambda^j}{|x|}
\ \ \ \ \ \ \ \ \ \ \ \ \ \ \ \ \ \ \ \ \ 
\label{g_trns}
\end{equation}
where $\lambda^0$ is the identity matrix and $a_{\mu}$ is a unit
vector along the direction $\mu$. Provided that the instanton
center is not on a lattice site, this transformation is well defined
everywhere.  After the transformation, all links lying entirely
outside $r_{\rm max}$ are equal to the unit matrix so the configuration
exactly obeys the usual periodic boundary conditions.
This instanton field is implemented on an $8^4$ lattice with
$\rho_0=10, \ r_{\rm max} = 3$ and on a $16^4$ lattice with
$\rho_0=20, \ r_{\rm max} = 7$.  Thus, upon moving from the smaller to 
the larger lattice, we have reduced the lattice spacing, measured in
units of the instanton radius, by nearly a factor of two.  
In order to study the robustness of
zero modes, random fluctuations with a given amplitude $\zeta$ are
superimposed. In particular, at each link a different SU(3) matrix is
constructed by exponentiating a linear combination of Gell-Mann
matrices with random coefficients in the range $[-\zeta, \zeta]$. The
gauge field at each link is then multiplied by these matrices.

The operator used to study the effects of the winding of the gauge
field background on the fermion sector is the chiral condensate
$\qbarq$ calculated on that background.  For staggered fermions
$\qbarq = {1 \over 3 V} Tr[D^{-1}]$ where $D$ is the
standard staggered Dirac operator and $V$ is the 
four-dimensional volume.  For domain wall fermions 
$\qbarq = {1 \over 12 V} Tr[(D^{-1})_{4d}]$
where $D$ is the five-dimensional Dirac operator of
\cite{Furman_Shamir}, $(D^{-1})_{4d}$ is its inverse with the fifth
dimension indices fixed so that it corresponds to a propagator between
four-dimensional quark fields that are projections of the
5-dimensional fields as prescribed in
\cite{Furman_Shamir}, and V is the four-dimensional volume. 
In particular, a four-dimensional Dirac spinor field
is formed by combining the two upper spin components of the
five-dimensional Dirac spinor field at $s=0$ with the lower two spin
components at $s=L_s-1$.  This Dirac operator contains an explicit
mass term that mixes the right and left chiralities with strength
$m_f$.  Antiperiodic boundary conditions along the time direction were
implemented for both staggered and domain wall fermions.  
The inversion was done using the conjugate gradient algorithm.  The
stopping condition (ratio of the residual over the norm of the source)
was set to: $10^{-5}$ for masses in the range $[10^{-1}, 10^{-2}]$,
$10^{-6}$ for masses in the range $(10^{-2}, 10^{-3}]$, $10^{-7}$ for
masses in the range $(10^{-3}, 10^{-4}]$, and $10^{-8}$ for masses in
the range $(10^{-4}, 10^{-5}]$.

The trace is over space-time, spin and color. A stochastic estimator
was used to calculate the trace.  To get reasonable estimates of the
average and error one would have to use a large number ($\sim 50$) of
Gaussian random vectors. However, in this paper the interest  is not
so much in the actual value of the trace since it is only used as a device
in studying the smallest eigenvalue (topological zero mode) of the
domain wall fermion Dirac operator.  To the extent that the Dirac
propagator which enters $\qbarq$ is dominated by this single
eigenvector of interest, the complete trace might be replaced by a
single diagonal element taken in a random direction.  This would give
the desired trace, multiplied by the overlap between the random vector
and the dominant zero mode.  If the same random vector is used for all
values of $m_0$, $m_f$ and $L_s$, it is expected that this overlap
will be essentially constant and the variations seen will be those
present in the full trace[6]. This strategy is followed in the
calculation.  However, in order to reduce the chance that the
contribution of the single eigenvector of interest is accidentally
suppressed by an unfortunate choice of random diagonal element, the
same set of ten Gaussian random vectors is used to approximate the
trace for all values of $m_0$, $m_f$ and $L_s$.  The error bars shown
in the figures represent the fluctuations seen among these ten random
vectors.  Given this small number of samples, it is not certain that
these error estimates are accurate.  However, additional calculation
suggests that this particular sample probably underestimates the
errors by less than a factor of two.

At infinite $L_s$ the explicit overlap formula for this formulation was derived in
\cite{Furman_Shamir}. Using that it can be shown that $\qbarq$
diverges as $\qbarq \sim |\nu| / m_f$, where $\nu$ is the index of the
Dirac operator as defined by the overlap formalism.  At finite $L_s$
the two chiralities do not completely decouple and therefore there is
a residual mass.  In free field theory this mass decreases
exponentially with $L_s$ and the effective quark mass is proportional
to the sum of the residual mass and the explicit mass $m_f$. In
particular \cite{PMV}:
\begin{equation}
m_{\rm eff} = m_0 (2- m_0) [m_f + (1-m_0)^{L_s}]
\label{mass}
\end{equation}
where $m_0$ is the mass (domain wall height) of the five dimensional theory.
In free field theory one flavor physics is obtained for
$m_0$ in the interval $[0,2)$ \cite{Kaplan}. In the interacting theory
the boundaries of this interval will be renormalized.
Assuming a similar modification of the effective quark mass for the
case of the instanton-like background one would expect that at finite
$L_s$, $\qbarq$ would behave as $\qbarq \sim |\nu| / [m_f + (1-
m_0^\prime )^{L_s}]$, where $m_0^\prime = m_0^\prime(m_0)$ is a
``renormalized'' domain wall height.

As mentioned above, the index of the Dirac operator in the overlap
formalism is naturally integer valued.  A method to measure this index
was developed in \cite{NN1} and used in \cite{NV}, \cite{EHN1},
\cite{EHN2}, \cite{EHN3}. The index is half the difference of the number $N_+$ of
positive minus the number $N_-$ of negative eigenvalues of the
Hamiltonian $H$ associated with the transfer matrix $T$.  This number
is the same as half the difference of the number of positive and negative
eigenvalues of the operator $D = \gamma_5 D_w$ where $D_w$ is the
standard Wilson Dirac operator evaluated at a mass which is the 
negative of the domain wall height, {\it i.e.} $-m_0$
\cite{NN1} \cite{Furman_Shamir}. 
For $m_0 < 0$ it can be shown that $N_+ = N_-$ for any background
gauge field. Therefore, by monitoring a few of the small positive and
negative eigenvalues of $D(m_0)$ as $m_0$ is varied between zero and
the positive value of interest, one can determine the number of
positive eigenvalues that crossed zero and became negative and the
number of negative eigenvalues that crossed zero and became
positive. The difference of the number of the two types of crossings
is the index.

For the classical backgrounds that we studied we found that on the
$8^4$ lattice the index changed from zero to one at $m_0 \approx 0.28$
and back from one to zero at $m_0 \approx 2.14$.  For the $16^4$ lattice
the index changed at $m_0 \approx 0.05$ and $m_0 \approx 2.01$. 
( Note this approach
to the free-field values of 0 and 2 is expected as one goes to the 
smaller lattice spacing implied by our $16^4$ instanton-like 
configuration. )
These values change by less than $1 \%$ when noise was added.
Therefore, in these intervals one expects $\qbarq$ to diverge as
$m_{\rm eff}$ is made small. At the crossing points the transfer
matrix has a unit eigenvalue and even at $L_s = \infty$ the two
chiralities do not decouple \cite{NN1}. Therefore, near a crossing
larger values of $L_s$ may be needed to see the expected $1/m$ behavior
\cite{Furman_Shamir}.

A ``sketch'' of the expected behavior of $\qbarq$ versus $m_f$ in the
presence of a gauge field background with net winding
is presented in Figure $1$.  In order to facilitate the comparison
with staggered fermions, we will also use $m_f$ for 
the usual staggered fermion mass. It is useful to analyze the mass 
dependence of $\qbarq$ by reference to the functional form of the 
spectral formula
\begin{equation}
\qbarq = [m / V] \sum_\lambda \{1 / (\lambda^2 + m^2)\}.
\label{spectral-form}
\end{equation}
This continuum equation is exact for the case of staggered fermions and
offers a useful framework for discussion of domain wall fermions.  We
can distinguish four distinct regions suggested by this functional
form.  In the large $m_f$ region (I), we expect the propagator to be
dominated by the mass term and therefore $\qbarq \sim 1/m_f$.  This
behavior is expected and seen for both domain wall and staggered
fermions.  Next, we define region (III) for staggered fermions, as the
mass range within which $m_f$ is small but larger than the smallest
eigenvalue $\lambda_{\rm min}$.  Here, $\lambda_{\rm min}$ is
presumably shifted away from zero by finite lattice spacing effects.
For staggered fermions $\lambda_{\rm min}$ is four-fold degenerate or
near degenerate for the case where noise has been added.  Region (III) is
defined similarly for domain wall fermions, except the condition $m_f
\ge \lambda_{\rm min}$ is replaced by $m_f \ge m_{\rm res}$, the
residual mass due to the mixing of chiralities between the walls.
(For the free field case, this is the $(1-m_0)^{L_s}$ term in
equation \ref{mass}.)  Although, domain wall fermions
in the $L_s\rightarrow \infty$ limit have $\lambda_{\rm min}=0$ 
its effects at finite $L_s$ are cut off by  this residual mass. Thus, for both
staggered and domain wall fermions, one expects the small eigenvalue
mode(s) to dominate the value of $\qbarq$ in region (III) and therefore
$\qbarq \sim 1/ (m_f V)$ where $V$ is the lattice volume. Region (II)
is the crossover region between region (I) and region (III).  In much
of this region the mass is much larger than $\lambda_{\rm min}$ but is
small enough to be relatively unimportant when compared with the rest
of the eigenvalue spectrum. Therefore, referring to equation $4$, 
one would expect $\qbarq \sim m_f$ in much of region (II).

The expected behavior in region (IV) is different for staggered 
and domain wall fermions.  For staggered fermions (solid line) 
$\lambda_{\rm min}$ is small but not zero. Thus, in region (IV), where 
$m_f < \lambda_{\rm min}$, the effects of the factor of $m_f$ in 
the numerator of eq. \ref{spectral-form} will dominate and
therefore one would expect $\qbarq \sim m_f$.  In contrast, for 
domain wall fermions (dashed line) and $m_f < m_{\rm res}$, we expect 
$m_{\rm res}$ to play the role of $m$ in eq. \ref{spectral-form} so that
in region (IV) $\qbarq \approx {\rm const.}$  For domain wall fermions,
region (IV) is dominated by the finite $L_s$, residual mixing of the 
two four-dimensional boundaries (see eq. \ref{mass}). 
In the numerical results of this paper
region (I) and part of region (II) are not present because the largest
mass is $m_f = 10^{-1}$. The focus is on region (III) where the
divergent $1/m_f$ behavior is expected and on the beginning of region
(IV) whose onset signals the need for larger values of $L_s$.

The numerical results are presented in Figures $2-8$. In Figure $2$
the staggered $\qbarq$ is plotted versus the quark mass $m_f$ in the
presence of the compactified, singular-gauge instanton background. Two
lattice volumes and three different noise amplitudes, $\zeta = 0,
0.01, 0.1$, are shown. For the case of the $8^4$ lattice with no
noise, one may be able to recognize divergent, $1/m_f$ 
behavior in $\qbarq$ for $10^{-2} \leq m_f \leq 10^{-1}$ indicating 
the presence of a near-zero mode.  This
divergent behavior does not extend to smaller $m_f$ presumably because
of the zero mode shift effect \cite{Smit_Vink}. As can be seen when
the lattice size is increased from $8^4$ to $16^4$ the instanton field
becomes ``smoother'' (lattice spacing is reduced by a factor of two)
and the divergent behavior becomes more pronounced now extending to
the region $10^{-4} \leq m_f \leq 10^{-2}$. For a detailed analysis the
reader is referred to \cite{Kaehler}.  However, when noise is added
the divergent behavior begins to disappear.  At noise amplitudes of
$0.1$ the divergent behavior is not present in the $8^4$ lattice while in the $16^4$
lattice it has been significantly reduced.

Figure $3$ is the same as Figure $2$ but now for the 
domain wall fermion $\qbarq$ with
$m_0=1.2$. Results are shown for the two volumes $8^4$ and $16^4$ and
for $L_s = 4, 6, 8, 10$. As can be seen, $\qbarq \sim 1/m_f$ for small
$m_f$. As expected, when $m_f$ becomes smaller than some value, the residual mixing
between the two chiralities becomes the dominant contribution to
$m_{\rm eff}$ and $\qbarq$ stops changing.  This value is the border between regions
(IV) and (III) sketched in Figure 1.  For the $16^4$ lattice and
for $L_s = 6, 8, 10$ one finds that $\qbarq \sim 1/m_f$ for $m_f$ as
small as $\approx 10^{-5}$.  For $L_s=4$ there is no signal of a
divergence because the effective quark mass is dominated by the finite
$L_s$ residual mass. For the $8^4$ lattice similar behavior is
observed but now for $L_s=8,10$. For $L_s=6$ a divergence is observed
but only for $m_f$ down to $ \approx 10^{-3}$.  When noise is added
with amplitudes $\zeta = 0.01$ and $\zeta = 0.1$ the behavior remains
unaffected for all practical purposes indicating robustness under high
frequency noise. For $\zeta=1$ the zero modes disappear (not shown
here) but this level of noise is so large that it presumably destroys the
winding. In particular, the index of the Dirac operator, as defined by the
overlap formalism, was found to be zero.

In Figure $4$ the $16^4$ lattice of Figure $3$ for $\zeta=0.1$
and $L_s=10$ is plotted again, this time with a $\qbarq = c_{-1}/m_f$ fit
for $10^{-5} \leq m_f \leq 10^{-3}$ superimposed. The coefficient 
$c_{-1}$ is $1.09(7) \times 10^{-6}$ and the $\chi^2$ per degree 
of freedom is $\approx 0.5$, demonstrating quantitatively the expected 
$m_f^{-1}$ behavior.

In order to study the $m_0$ dependence, $\qbarq$ versus $m_0$ is plotted in
Figure $5$ for fixed $m_f= 5 \times 10^{-4}$. The vertical lines indicate the
values of $m_0$ where $\gamma_5 D_w(m_0)$ has a crossing.  One can see that near
the crossing points larger values of $L_s$ are required before
$\qbarq$ becomes independent of $L_s$. However, there is a large range
of $m_0$ in between the crossings for which $\qbarq$ does not change
much when $L_s$ is changed from $8$ to $10$.  This means that no fine
tuning of $m_0$ is needed even when large amounts of ultraviolet noise
are added.  The shape of the curves can be attributed to the wave
function normalization factor which is a function of $m_0$. In the
free theory this factor is $m_0 (2-m_0)$
\cite{PMV}.
	 
To further verify robustness under changes of $m_0$, $\qbarq$ versus $m_f$
is studied for $m_0=0.75$. As can be seen from Figure $5$ this value
of $m_0$ is in the onset of the region where the $L_s$ dependence
becomes stronger. The results for $L_s=4,6,8,10,12$ are presented
in Figure $6$. Also, $L_s=24$ is presented for the $8^4$ lattice.  As
can be seen, the zero mode effects are maintained and are robust under
high frequency noise but larger $L_s$ is
needed before $\qbarq$ becomes independent of $L_s$. For the $16^4$ lattice
$\qbarq$ becomes independent of $L_s$ for $L_s=12$, but for the $8^4$ lattice
this does not happen until $L_s=24$. This is expected since the
crossing point for the $8^4$ lattice is closer to $m_0=0.75$ than the crossing
point for the $16^4$ lattice.

If $m_0$ is allowed to get close to the crossing point one
would expect that in order to maintain the zero mode effects much
larger values of $L_s$ may be needed. Figure $7$ is the same as Figure
$6$, except that here $m_0$ is chosen a fixed distance of $0.2$ from the crossing
point. For the $8^4$ lattice $m_0 = 0.5$ and for the $16^4$ lattice
$m_0=0.25$.  The plots for $L_s = 8, 12, 16$ and $ 24$ are shown. Also
for the $8^4$ lattice the plot for $L_s=48$ is presented. As can be seen for
both volumes $L_s=24$ is barely adequate to maintain the zero mode
effect for masses $5 \times 10^{-3} < m_f$. In order to maintain zero
mode effects for masses $10^{-4} < m_f < 5 \times 10^{-3}$ an $L_s=48$
is needed for the $8^4$ lattice.

Finally, $\qbarq$ is plotted versus $L_s$ in figure $8$ for zero noise
amplitude, fixed $m_f = 5 \times 10^{-4}$ and for three values of $m_0
= 0.5, 0.75$ and $1.2$ corresponding to circles, squares and
crosses. The fits are to a function of the form $1 / [ c_0 + c_1
e^{-c_2 L_s}]$. For the three values of $m_0$ the fitting range of
$L_s$ is $[8,48]$, $[6,48]$ and $[4,48]$, the coefficient $c_0 =
0.183(8), 0.44(3)$ and $0.91(31)$, and the $\chi^2/{\rm dof}$ is $4
\times 10^{-2}$, $4 \times 10^{-2}$ and $2 \times 10^{-4}$.  "(Note,
these fits were performed by minimizing the simplest $\chi^2$, which
did not incorporate the strong correlation between the fluctuations at
different values of $L_s$.  The presence of such correlations,
resulting from using the same random source vectors for each $L_s$, is
reflected in the abnormally small values of $\chi^2$.)  Again, the
quality of these fits demonstrates the expected $1/ m_{\rm eff} \sim 1
/ [m_f + m_{\rm res}]$ dependence of $\qbarq$ but this time $m_f$ is
held fixed while $m_{\rm res} \sim e^{-c_2 L_s}$ is varied. As
expected, the decay rate is fast for $m_0=1.2$ but as $m_0$ gets
closer to the value where $\gamma_5 D_w$ has a crossing the decay rate
becomes slower.  Also, as expected from the discussion relating to
equation $4$ and figure $1$ when $m_{\rm res}$ becomes large the
behavior changes from the monotonic behavior of regions IV and III to
that of region II. For $m_0=0.5$ this can be seen for $L_s=4, 6$ and
for $m_0=0.75$ it can be seen for $L_s=4$. For $m_0=1.2$ presumably
$m_{\rm res}$ does not get large even for $L_s=4$ and there is no
change in the monotonic behavior.

\section{Conclusions}

In conclusion, for a classical, instanton-like background the domain
wall fermion chiral condensate diverges as $\qbarq \sim 1/ m_f$ for
$m_f$ as small as $10^{-5}$ and $L_s$ as small as $\sim 10$ lattice
spacings.  This behavior was observed on lattices of size $8^4$ and
$16^4$ and was robust under random, high frequency noise with
amplitudes as large as $\zeta = 0.1$.  This $1/m_f$ divergence was
observed for a wide range of values of $m_0$, indicating that there is
no need for a fine tuning of $m_0$.  Furthermore, at fixed $m_f$ and
$m_0$ the chiral condensate approaches its $L_s=\infty$ asymptotic
value exponentially fast with a rate that becomes faster as $m_0$ is
varied away from the point where the index of the Dirac operator
changes.  Therefore, domain wall fermions continue to show their
spectacular $L_s=\infty$ zero-mode properties even for values of $m_f$
and $L_s$ that are practically accessible to contemporary numerical
simulations.  Numerical simulations of QCD are currently under way
\cite{DWF_Columbia} to study the zero mode effects of domain wall
fermions on gauge field backgrounds with realistic quantum fluctuations.


\section*{Acknowledgments}

The numerical calculations were done on the 400 Gflop QCDSP computer at
Columbia University. We would like to thank R. Edwards and R. Narayanan
for helpful discussions. This research was supported in part by the DOE
under grant \# DE-FG02-92ER40699.


\vfill

%
%
\if \epsfpreprint Y

\eject

\figure{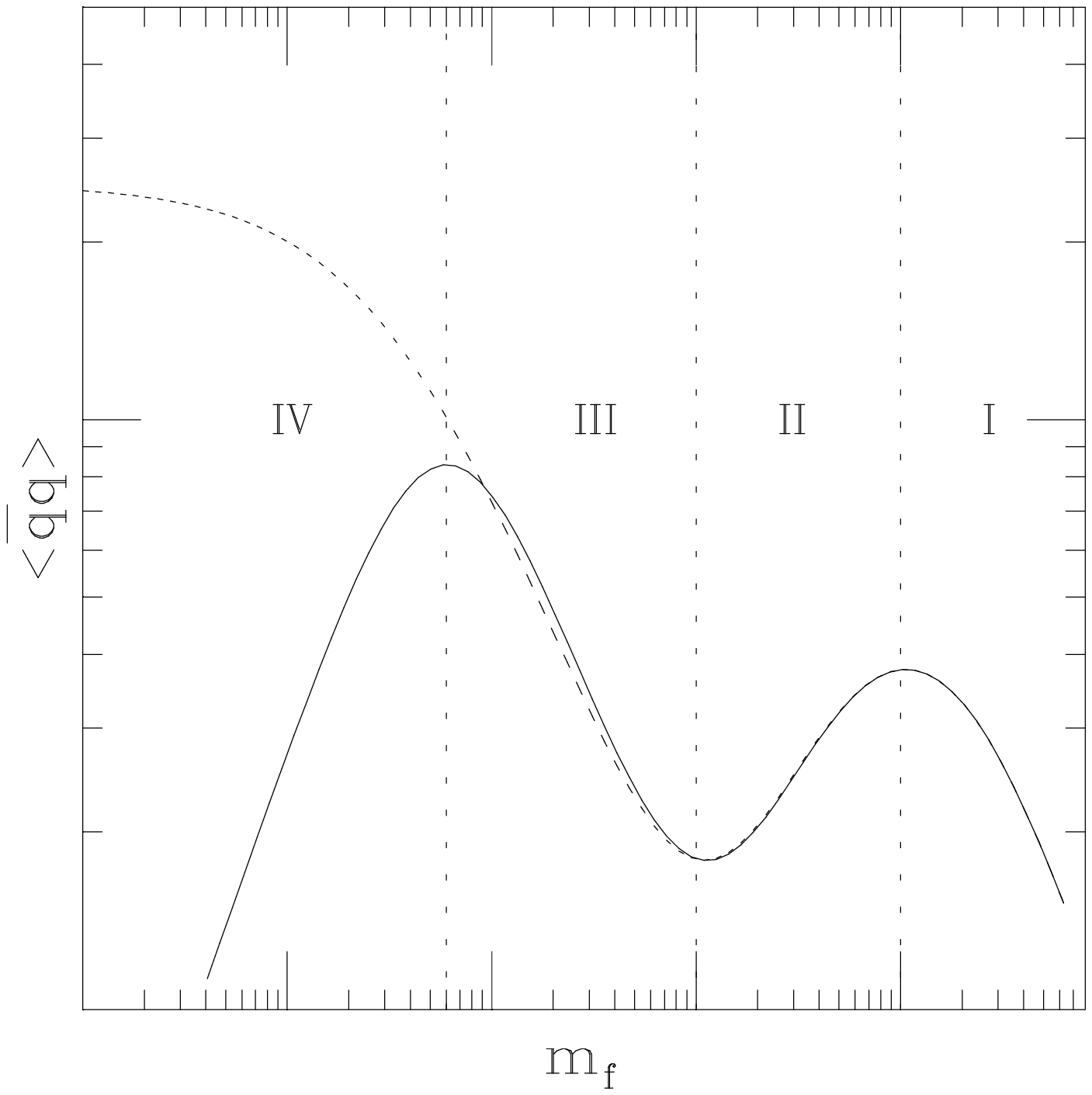}{\one}{\figsize}

\figure{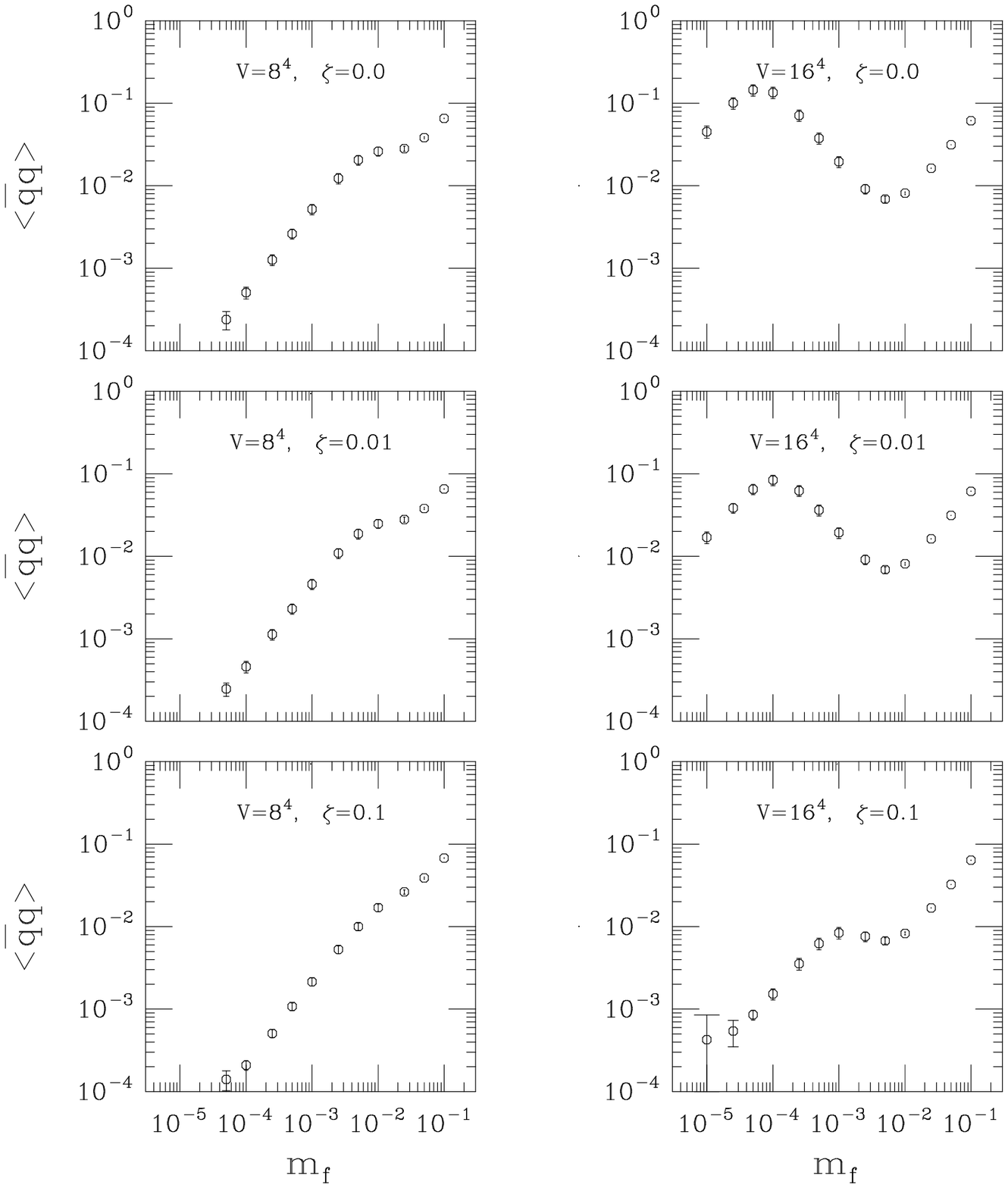}{\two}{\figsize}

\figure{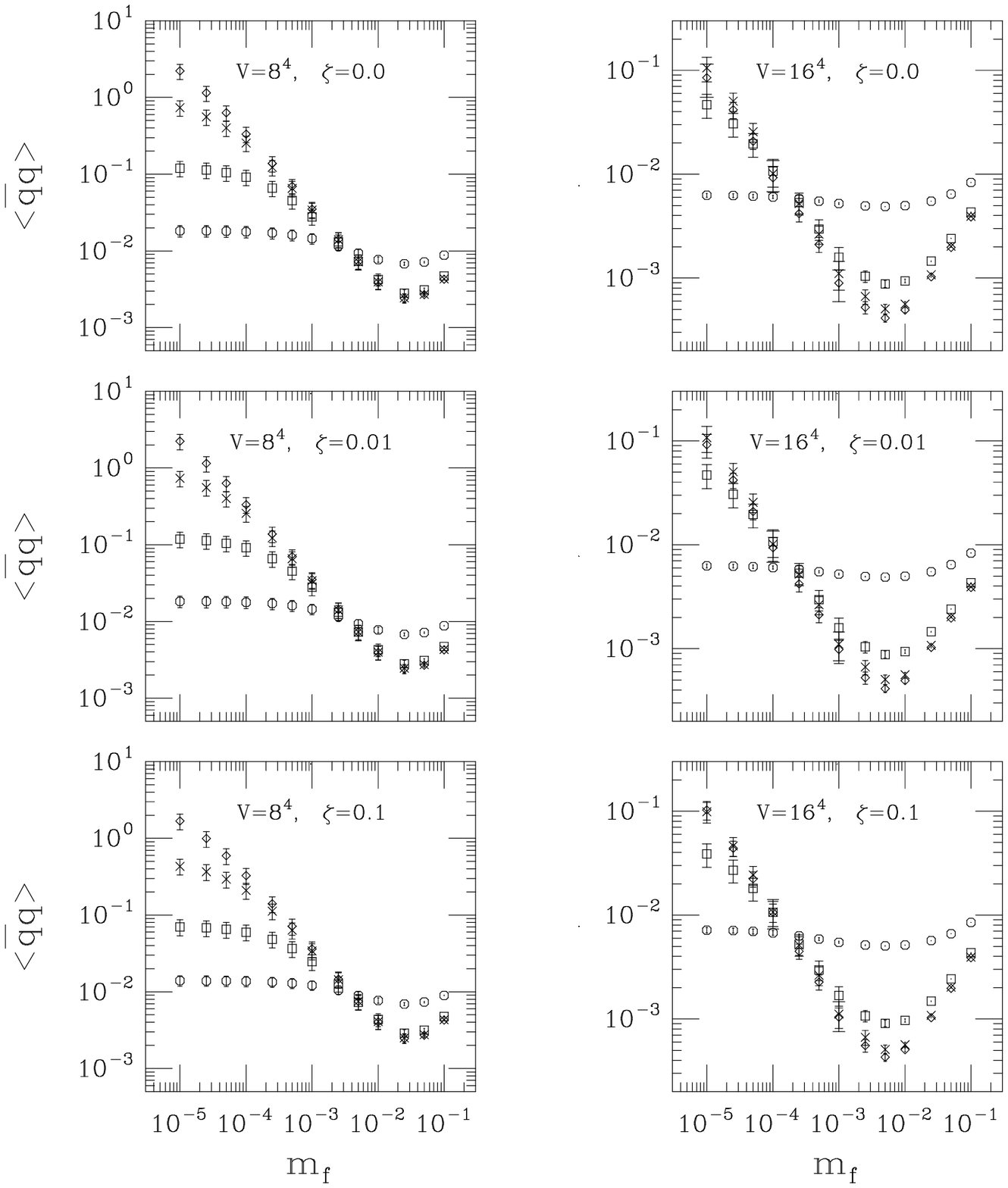}{\three}{\figsize}

\figure{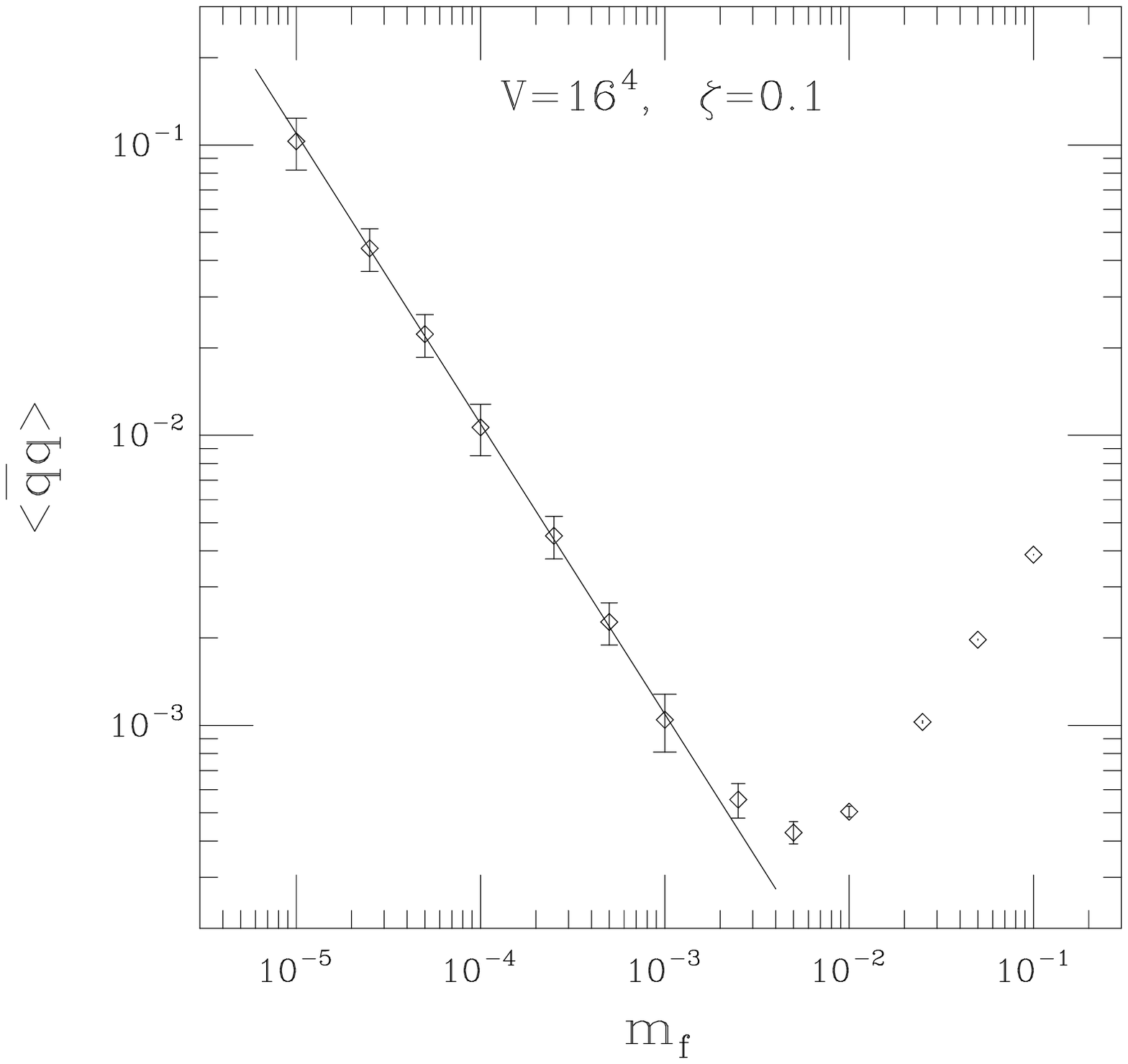}{\four}{\figsize}

\figure{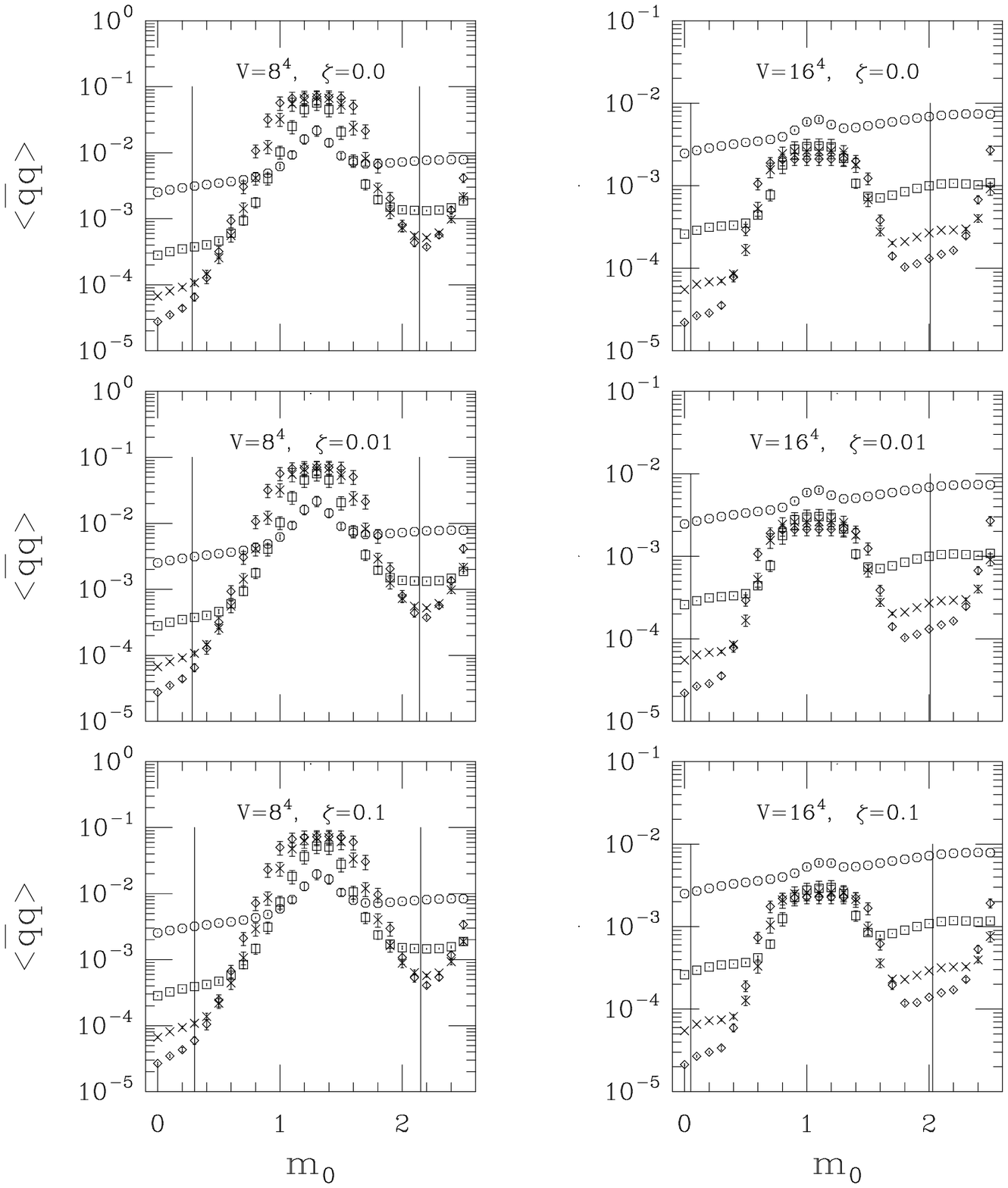}{\five}{\figsize}

\figure{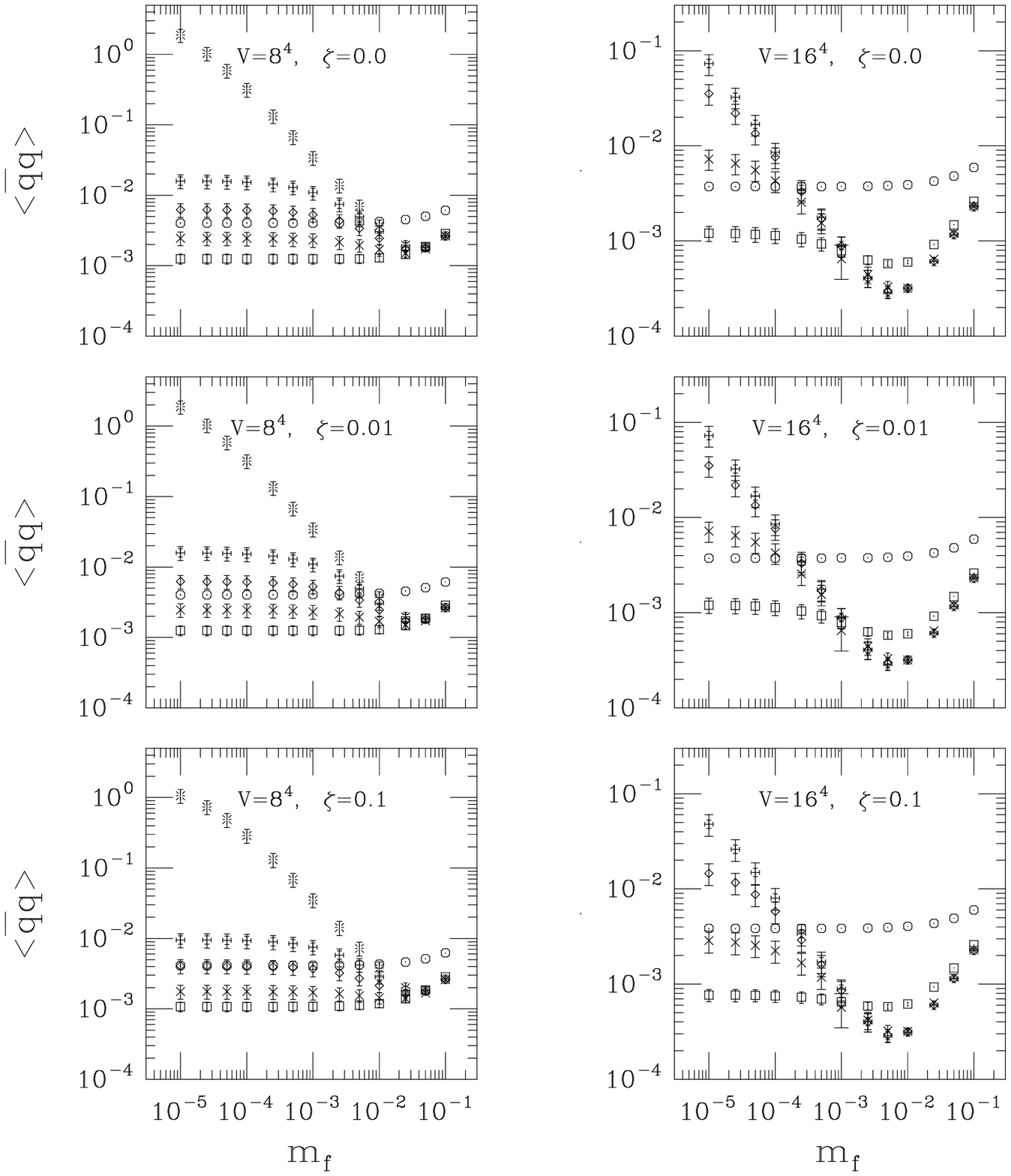}{\six}{\figsize}

\figure{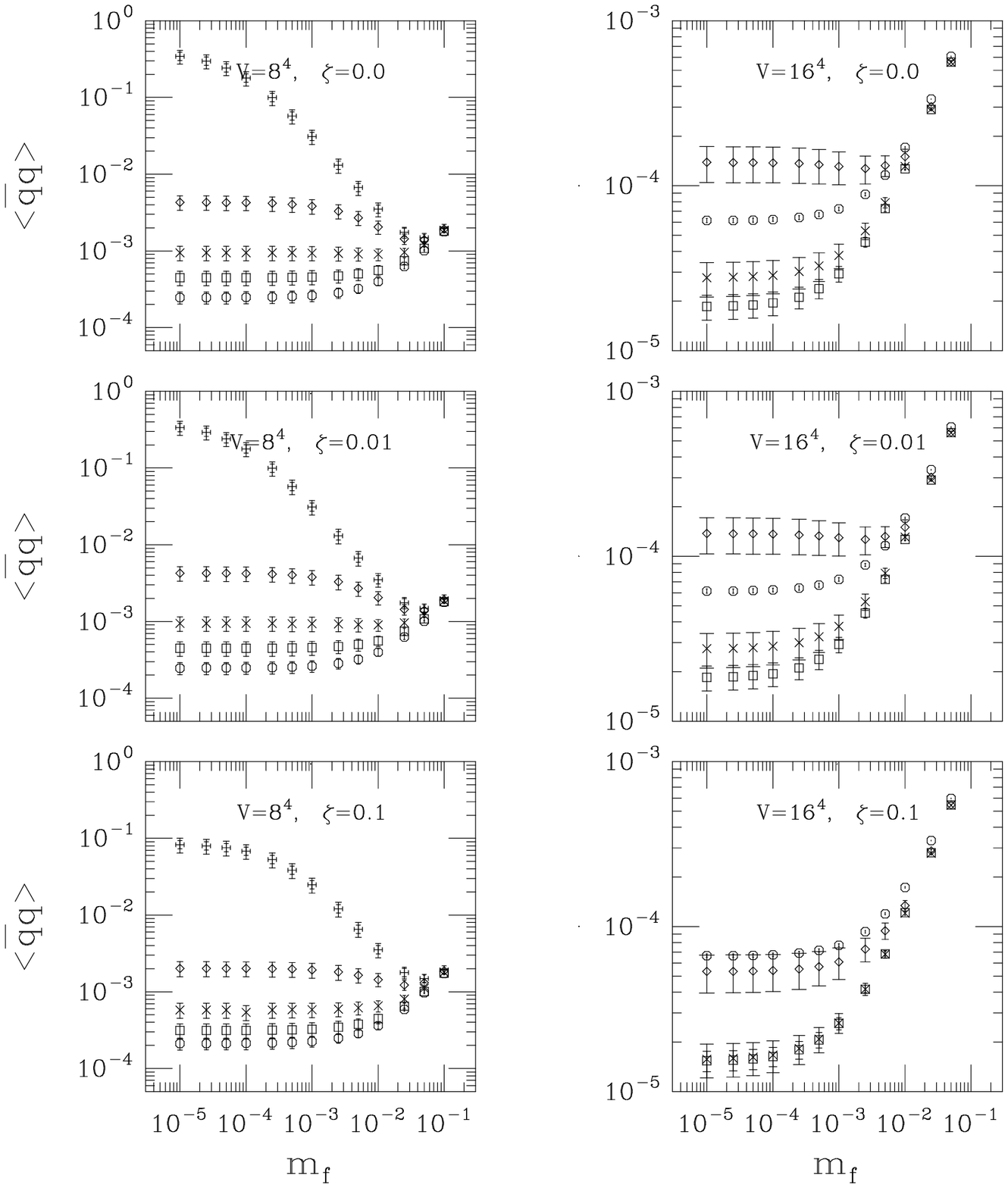}{\seven}{\figsize}

\figure{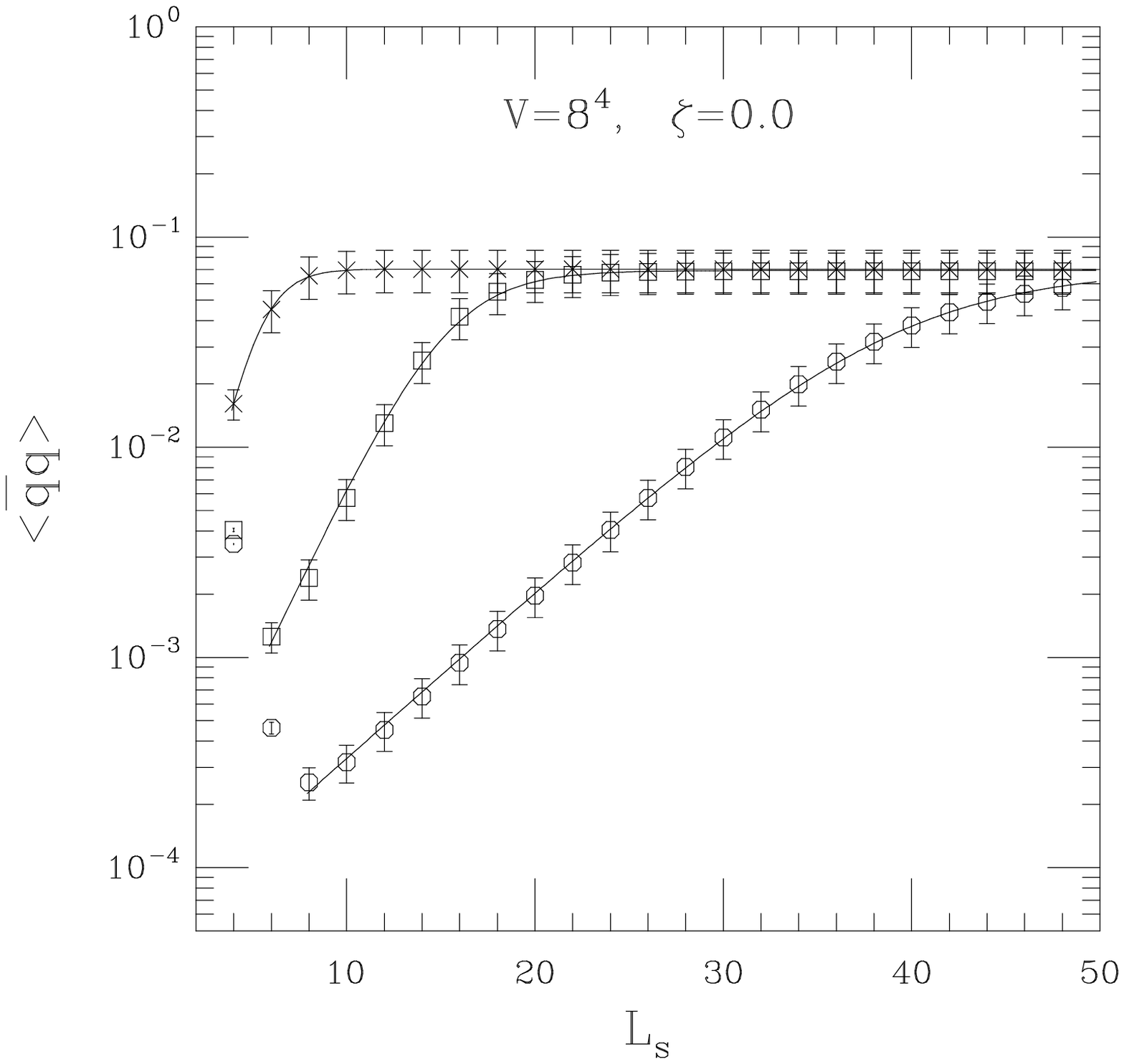}{\eight}{\figsize}

\fi

\if \epsfpreprint N 

\eject

\section* {Figure Captions.}

\noindent{\bf Figure 1:} \one

\noindent{\bf Figure 2:} \two

\noindent{\bf Figure 3:} \three

\noindent{\bf Figure 4:} \four

\noindent{\bf Figure 5:} \five

\noindent{\bf Figure 6:} \six

\noindent{\bf Figure 7:} \seven

\noindent{\bf Figure 8:} \eight

\fi

\end{document}